\documentclass[%
 article,twocolumn,
 superscriptaddress,
 amsmath,amssymb,
aps,
prb,
citeautoscript
]{revtex4-2}
\usepackage[normalem]{ulem}
\usepackage{natbib}
\usepackage{xcolor}
\usepackage{graphicx}
\usepackage{amsmath}
\usepackage{physics}
\usepackage[colorlinks=true,allcolors=blue]{hyperref}%
\DeclareUnicodeCharacter{2212}{-}

\usepackage{xspace}

\renewcommand{\sout}[1]{\unskip}
\newcommand{\figref}[2]{\hyperref[#1]{Figure~\ref{#1}\textbf{#2}}}
\newcommand{\figureref}[2]{\hyperref[#1]{Figure~\ref{#1}\textbf{#2}}}

\usepackage{nameref}
\usepackage{hyperref}
\hypersetup{colorlinks,allcolors=blue}

\begin{document}

\title{Collective Magnetic Excitations in a Photo-excited Electron-doped Cuprate Superconductor} 
\date{\today}


\author{Daniel~Jost}
\thanks{D.J. and J.L. contributed equally to this work.}
\affiliation{Stanford Institute for Materials and Energy Sciences, SLAC National Accelerator Laboratory, Menlo Park, CA, USA}

\author{Jiarui~Li}
\thanks{D.J. and J.L. contributed equally to this work.}
\affiliation{Stanford Institute for Materials and Energy Sciences, SLAC National Accelerator Laboratory, Menlo Park, CA, USA}

\author{Jordyn Hales}
\affiliation{Department of Chemistry, Emory University, Atlanta, GA 30322, United States}

\author{Jonathan~Sobota}
\affiliation{Stanford Institute for Materials and Energy Sciences, SLAC National Accelerator Laboratory, Menlo Park, CA, USA}

\author{Giacomo~Merzoni}
\affiliation{European XFEL, Holzkoppel 4, Schenefeld, 22869, Germany}
\affiliation{Dipartimento di Fisica, Politecnico di Milano, piazza Leonardo da Vinci 32, I-20133 Milano, Italy}

\author{Leonardo~Martinelli}
\affiliation{Dipartimento di Fisica, Politecnico di Milano, piazza Leonardo da Vinci 32, I-20133 Milano, Italy}

\author{Shuhan Ding}
\affiliation{Department of Chemistry, Emory University, Atlanta, GA 30322, United States}

\author{Kejun~Xu}
\affiliation{Department of Applied Physics, Stanford University, Stanford, CA, USA}

\author{Justine~Schlappa}
\affiliation{European XFEL, Holzkoppel 4, Schenefeld, 22869, Germany}

\author{Andreas~Scherz}
\affiliation{European XFEL, Holzkoppel 4, Schenefeld, 22869, Germany}

\author{Robert~Carley}
\affiliation{European XFEL, Holzkoppel 4, Schenefeld, 22869, Germany}

\author{Benjamin~E.~Van~Kuiken}
\affiliation{European XFEL, Holzkoppel 4, Schenefeld, 22869, Germany}

\author{Teguh~C.~Asmara}
\affiliation{European XFEL, Holzkoppel 4, Schenefeld, 22869, Germany}

\author{Le~Phuong~Hoang}
\affiliation{European XFEL, Holzkoppel 4, Schenefeld, 22869, Germany}

\author{Laurent~Mercadier}
\affiliation{European XFEL, Holzkoppel 4, Schenefeld, 22869, Germany}

\author{Sergii Parchenko}
\affiliation{European XFEL, Holzkoppel 4, Schenefeld, 22869, Germany}

\author{Martin~Teichmann}
\affiliation{European XFEL, Holzkoppel 4, Schenefeld, 22869, Germany}

\author{Patrick~S.~Kirchmann}
\affiliation{Stanford Institute for Materials and Energy Sciences, SLAC National Accelerator Laboratory, Menlo Park, CA, USA}

\author{Giacomo~Ghiringhelli}
\affiliation{Dipartimento di Fisica, Politecnico di Milano, piazza Leonardo da Vinci 32, I-20133 Milano, Italy}
\affiliation{CNR-SPIN, Dipartimento di Fisica, Politecnico di Milano, I-20133 Milano, Italy}


\author{Brian~Moritz}
\affiliation{Stanford Institute for Materials and Energy Sciences, SLAC National Accelerator Laboratory, Menlo Park, CA, USA}

\author{Zhi-Xun~Shen}
\affiliation{Stanford Institute for Materials and Energy Sciences, SLAC National Accelerator Laboratory, Menlo Park, CA, USA}
\affiliation{Geballe Laboratory for Advanced Materials, Stanford University, Stanford, CA, USA}
\affiliation{Department of Applied Physics, Stanford University, Stanford, CA, USA}
\affiliation{Department of Physics, Stanford University, Stanford, CA, USA}

\author{Thomas~P.~Devereaux}
\affiliation{Stanford Institute for Materials and Energy Sciences, SLAC National Accelerator Laboratory, Menlo Park, CA, USA}
\affiliation{Department of Materials Science and Engineering, Stanford University, Stanford, CA, USA}
\affiliation{Geballe Laboratory for Advanced Materials, Stanford University, Stanford, CA, USA}

\author{Yao~Wang}
\email{yao.wang@emory.edu}
\affiliation{Department of Chemistry, Emory University, Atlanta, GA 30322, United States}

\author{Wei-Sheng Lee}
\email{leews@stanford.edu}
\affiliation{Stanford Institute for Materials and Energy Sciences, SLAC National Accelerator Laboratory, Menlo Park, CA, USA}

\date{\today}




\begin{abstract}
Elucidating the microscopic behavior of cuprates under ultrafast photoexcitation offers critical insights into their highly correlated out-of-equilibrium states. Although quasiparticle dynamics have been investigated extensively, the behavior of collective magnetic excitations remains comparatively unexplored. Here, we use time-resolved resonant inelastic X-ray scattering (trRIXS) at the Cu $L_3$-edge to track the collective magnetic excitations (paramagnons) in an optimally electron-doped cuprate driven out-of-equilibrium by a femtosecond pump laser pulse. Upon pumping, we observed an anti-Stokes signal associated with paramagnon generation, which modifies the paramagnon dispersion near the zone center, although the bandwidth remained unchanged. Moreover, the spectral weight exhibits a momentum-dependent variation across the Brillouin zone. The light-driven boost of the paramagnon population and the resulting spectral-weight transfer could provide new leverage to manipulate the properties of cuprates.

\end{abstract}
\maketitle

In high-$T_c$ superconducting cuprates, antiferromagnetic (AFM) correlations and doped charge carriers form a complex and strongly intertwined many-body wavefunction, which can manifest distinct quantum phases by changing doping and temperature\,\cite{Keimer2015}. By driving charge and spin degrees of freedom via photoexcitation, the wavefunction could be transiently modified, affecting the electronic properties of cuprates\,\cite{Basov2017} and potentially lead to a state without an equilibrium analog \cite{Wang2013,Wang2018,Torre2021}. Time-resolved optical spectroscopy\,\cite{Giannetti2016}, angle-resolved photoemission spectroscopy\,\cite{Zonno2021}, and X-ray scattering\,\cite{Mankowsky2014,Forst2014,Mitrano2019,Jang2022,Wandel2022} have elucidated extensive information about the photoexcited response of electronic bands, superconducting pairs, and the lattice, as well as charge order. However, collective excitations of spin, which bears the hallmark of spin exchange interaction and information of its coupling to the charges \cite{Keimer2015}, remain largely unexplored in photoexcited cuprates due to the lack of a suitable probe.

Time-resolved resonant inelastic X-ray scattering (trRIXS), which is capable of tracking collective excitations in time, energy, and momentum~\cite{Jost:2025,Mitrano2020} can be an ideal tool. Although the advent of X-ray free-electron lasers (FEL) a decade ago has enabled trRIXS measurements\,\cite{Dean2016,Mazzone2021,Mitrano2019,Lu2020,Paris2021}, comprehensive high-resolution trRIXS measurements on cuprates have only recently become possible due to the availability of high-repetition-rate soft X-ray FELs and high-resolution trRIXS instruments \cite{Schlappa2025}.

\begin{figure}[t]
    \centering
    \includegraphics[]{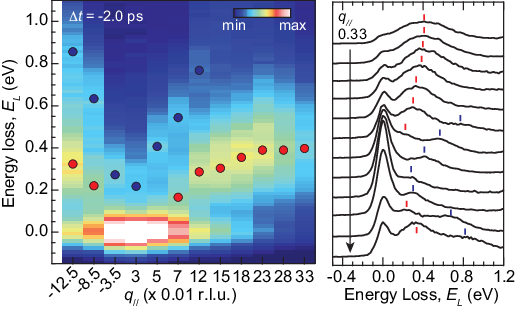}
    \caption{Energy-momentum RIXS intensity map (left) taken before time zero ($\Delta t$ = -2.0 ps). The waterfall plot of RIXS spectra are shown in the right panel. The red and blue markers indicate the paramagnon and plasmon peaks, respectively.}
    \label{fig:Equilibrium}
\end{figure}

\begin{figure*}[t]
    \centering
    \includegraphics[]{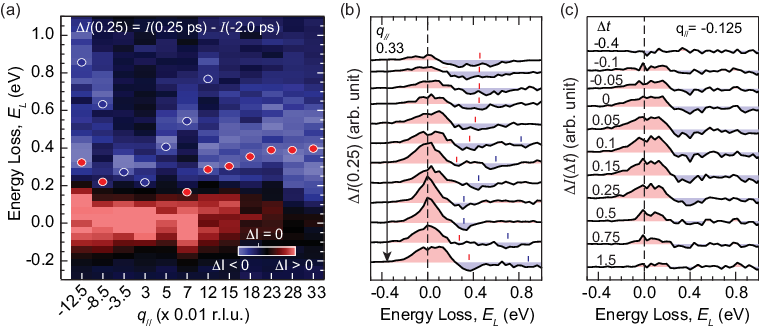}
    \caption{ (a) Differential RIXS intensity map obtained by subtracting the RIXS map taken at $\Delta t=0.25$ ps from that taken at -2.0\,ps. The red and blue markers indicated the peak positions of paramagnon and plasmon, same as those shown in Fig. \ref{fig:Equilibrium}. (b) Waterfall plot of the differential RIXS spectra of panel (a). The red and blue ticks  indicate the variations associated near the peaks of paramagnon and plasmon, respectively. (c) Differential spectral at $q_{\parallel} = -0.125$ \textit{r.l.u.} taken at different $\Delta t$.}
    \label{fig:differential_RIXS_data}
\end{figure*}

In this work, we focus on the electron-doped cuprate Nd$_{2-x}$Ce$_x$CuO$_4$ (NCCO), a material whose phase diagram is characterized by a prominent antiferromagnetic (AFM) phase \cite{Motoyama2007, Greene2020}. The underlying AFM correlation plays a critical role in shaping the electronic structure \cite{Armitage2010} and influences the emergence of superconducting coherence in the underdoped regime\cite{Xu2024}. At optimal doping ($x \sim 0.15$), the spin correlation length becomes short-ranged - about 20 unit cells \cite{Motoyama2007}- yet still couples to quasiparticles at the Fermi energy, reconstructing the Fermi surface and producing a pseudogap at the “hot spot” where it intersects the AFM zone boundary \cite{Armitage2010, Greene2020,He2019}. Interestingly, time-resolved angle-resolved photoemission spectroscopy (tr-ARPES) has demonstrated that photoexcitation results in the filling of this pseudogap, an effect previously linked to a decrease in the spin correlation length \cite{Boschini2020}. Nevertheless, a comprehensive understanding of the underlying microscopic dynamics remains elusive, as a direct spectroscopic interrogation of collective spin excitations in the photoexcited state is still lacking.

Here, we use trRIXS to track the time evolution of magnetic excitations in electron-doped superconducting cuprate Nd$_{2-x}$Ce$_x$CuO$_4$ (NCCO) at optimal doping $x \sim 0.15$. Further details of the experiment are described in Supplemental Material \cite{Sup}.  Although a pump laser pulse at 800 nm (corresponding to a photon energy of $\sim 1.55$ eV) is most commonly used for photoexcitation, we instead excited the system with 400 nm laser pulses ($\sim 3.1$ eV) at a fluence of 7\,mJ/cm$^2$, as the optical absorption is higher at this wavelength \cite{Onose2004}. We note that in optimally doped NCCO the optical conductivity is essentially featureless above 1.0 eV \cite{Onose2004}, implying that the excitation induced by the 400 nm pump pulse is similar to that produced by the 800 nm pump pulse. In both cases, photoexcitation creates ``hot" electrons, whose temporal evolution is often described phenomenologically using a multi-temperature model \cite{Perfetti2007, Zonno2021}. Our observation of collective magnetic excitations adds the missing element - the spin degree of freedom - to the microscopic picture of photoexcited cuprates.

Fig.~\ref{fig:Equilibrium} shows the RIXS intensity as a function of energy loss $E_L$ and momentum transfer $q_{\parallel}$ along the Cu–O bond direction (\textit{i.e.}, the $h$-direction), measured with X-ray FEL pulses before time zero, defined as the arrival time of the pump pulse at the sample. Dispersive paramagnons are evident, emanating from the zone center ($q_{\parallel}=0$) toward the zone boundary ($q_{\parallel}=0.5$) with a bandwidth of approximately 400 meV. In addition, another branch of rapidly dispersing charge modes, acoustic plasmons\,\cite{Hepting2018, Lin2020}, can be resolved near the zone center. These spectra are consistent with previous static RIXS measurements using X-rays from synchrotron light sources\,\cite{Lee2014,Ishii2014,Hepting2018,Lin2020}. 

\begin{figure*}[t]
    \centering
    \includegraphics[]{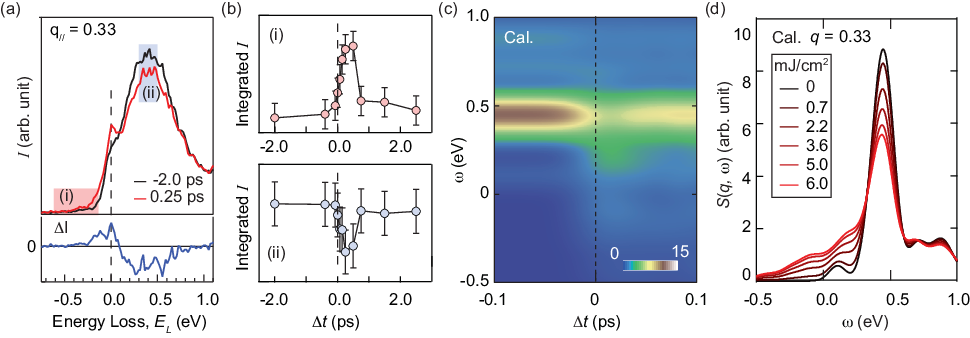}
    \caption{(a) RIXS spectra at $q_{\parallel}=0.33$ r.l.u. before and after time-zero with the differential spectrum shown in the lower panel. (b) Time traces of the integrated intensity within windows (i) and (ii) indicated in panel (a). The error bars were estimated from the noise level of the RIXS spectrum. (c) Time dependence of calculated dynamical spin structure factor $S(q,\omega,t)$ at $\mathbf{q}=(0.33,0)$ for 6\,mJ/cm$^2$ pump, obtained using the single-band Hubbard model. (d) Fluence dependence of calculated $S(q,\omega,t)$ at the center of the pump ($\Delta t = 0$) for same momentum as panel (c).}
    \label{fig:Paramagnon_q0p33}
\end{figure*}

To evaluate the light-driven effects, we analyzed the differential RIXS spectra $\Delta I$, as shown in Figs.~\ref{fig:differential_RIXS_data}(a,b), obtained by subtracting the spectra at time delay $\Delta t = 0.25$ ps from those at $\Delta t = -2.0$ ps. The changes, particularly the intensity depletion highlighted in blue, reflect modifications in the spectral line shapes of paramagnons and plasmons across the Brillouin zone. As will be elaborated below, these spectral changes extend into the quasi-elastic region, giving rise to an apparent increase in the quasi-elastic peak intensity, highlighted in red. As a function of time, shown in Fig.~\ref{fig:differential_RIXS_data}(c) for a representative momentum $q_{\parallel} = -0.125$ \textit{r.l.u.}, the light-induced response exhibits a rapid onset and attains its maximum amplitude at approximately 0.1 ps. This characteristic timescale is comparable to the experimental temporal resolution of $\sim$0.15 ps, indicating that the effect is predominantly generated during the pump pulse. Afterward, the light-induced changes diminish and have almost returned to their equilibrium form after 1.5 ps. While plasmons also show interesting variations, their three-dimensional nature requires out-of-plane momentum-dependent measurements for full characterization \cite{Hepting2018}, which is beyond the scope of our data. Therefore, in this letter, we focus only on the behavior of the paramagnons.

We first examine raw RIXS spectra taken at $q_{\parallel} = 0.33$ r.l.u., a momentum where only the paramagnon excitation is present. Upon pumping, as shown in Fig.~\ref{fig:Paramagnon_q0p33}(a), the paramagnon peak intensity is suppressed without changing the peak position. In addition, the spectral weight in the energy gain regime is increased at least up to $E_L$ = -0.3 eV,  indicating the presence of anti-Stokes scattering from certain modes, whose population substantially increases due to the pump. The modes, whose energy scales are notably larger than the maximal phonon energy in cuprates ($\sim$ 80 meV), are most likely paramagnons, given that they are the dominant mode at this momentum with a matching energy scale. Because recovery of the magnetic degree of freedom requires dissipation of pump-induced paramagnons, the energy-gain spectral weight should follow the same temporal evolution as the paramagnon peak in the energy-loss channel. This is confirmed in Fig. \ref{fig:Paramagnon_q0p33}(b): both the integrated energy-gain spectral weight (window (i)) and the paramagnon peak intensity (window (ii)) show nearly identical time dependence, supporting the assignment of the energy-gain features to a light-induced paramagnon population. We remark that due to the paramagnon's broad peak width (\textit{i.e.} short lifetime), anti-Stokes scattering does not manifest as a distinct peak, but rather as additional spectral weight on the energy gain side (see Supplemental Material Fig. S2 \cite{Sup}), as observed in our data. Also, in this situation, the spectral weight at zero energy becomes finite, resulting in an increase of the elastic peak intensity, as seen in our data (Figs. \ref{fig:differential_RIXS_data} and \ref{fig:Paramagnon_q0p33}). 

To address the observed light-driven paramagnon variation, we perform time-dependent exact diagonalization simulations for the time-resolved dynamical spin structure factor $S(\mathbf{q},\omega,\Delta t)$, using the single-band Hubbard model on a 12-site cluster with an electron doping of 16.7\%, which is the minimal achievable doping in our calculation (see Supplemental Material \cite{Sup}). Pump photon energy and polarization are identical to those used in the experiment. Figs.~\ref{fig:Paramagnon_q0p33} (c) and (d) present the $S(\mathbf{q},\omega,\Delta t)$ calculated at its smallest accessible nonzero momentum $q_\parallel$ = 0.33\,r.l.u.~for our cluster. In equilibrium, before the pump pulse arrives, the spectrum exhibits a peak at $400$\,meV corresponding to the paramagnon excitation in NCCO. At the center of the pump ($\Delta t=0$), this peak height decreases rapidly. At the same time, lower energy spectral weight extending to the energy gain ($\omega < 0$) increases, manifest as the broadening of the paramagnon excitation. We also found that the main peak position is essentially unchanged (Fig.~\ref{fig:Paramagnon_q0p33}(c)) and not sensitive to the pump fluence (Fig.~\ref{fig:Paramagnon_q0p33}(d)), which primarily impacts spectral weight suppression and anti-Stokes spectral weight generation. The behavior of $S(\mathbf{q},\omega,\Delta t)$ is consistent with our experimental observation at $q_\parallel$ = 0.33\,r.l.u.. We have also performed trRIXS calculations, which match the behaviors of the calculated $S(\mathbf{q},\omega,\Delta t)$, ruling out potential artifacts stemming from the finite-core-hole life time and matrix element effects (Supplemental Material Fig. S3). 

\begin{figure*}[t]
    \centering
    \includegraphics[]{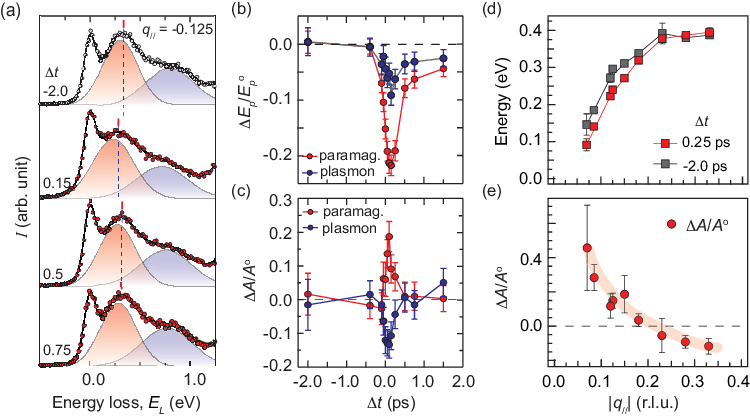}
    \caption{(a) RIXS spectra at representative time delays at $q_{\parallel} = -0.125$\,r.l.u.. The red ticks are the apparent paramagnon peak position in raw data the dashed lines are guide-of-the eye for the shift of the paramagnon peak position. The red and blue shaded peaks represent the fitted paramagnon and plasmon. (b) Time traces of normalized peak position changes for paramagnon and plasmon. (c) Time traces of normalized spectral weight changes for paramagnon and plasmon. (d) Summary of the fitted paramagnon positions E$_{p}$ before (black markers) and after (red markers) time zero. (e) Momentum dependence of the normalized change of the paramagnon spectral weight (area of the fitted paramagnon) $\Delta A = {A^{(0.25 \text{ps})}} - {A^o}$, where ${A^o} = {A^{(-2.0 \text{ps})}}$. Shaded area are guide-to-the-eye. The error bars were estimated via standard propagation of uncertainty from the 95\% confidence level from the fit of the peak parameter }
    \label{fig:q_dep}
\end{figure*}

Moving toward the Brillouin zone center, the behavior of the paramagnons with small $q_{\parallel}$ appears to be more complex. As shown in Fig. \ref{fig:q_dep}(a), at $q_{\parallel} = -0.125$ r.l.u., while broadened, the paramagnon peak position in the raw spectra shifts to lower energy near time zero due to the pump. To disentangle it from the evolution of the plasmon in the same spectrum, we fitted the trRIXS spectra to extract the paramagnon peak as a function of $\Delta t$. Given that the paramagnon is overdamped and the evolution of its anti-Stokes and Stokes part are locked near time-zero (Fig.~\ref{fig:Paramagnon_q0p33}(b)), we approximate the paramagnon spectrum with a single Gaussian peak spanning across zero energy, as superimposed in Fig. \ref{fig:q_dep}(a) (see also Supplementary Material and Fig. S4 \cite{Sup}).  As summarized in Fig. \ref{fig:q_dep}(b), the fitted paramagnon peak position softens by approximately 20\% , accompanied by an increase of paramagnon spectral weight (Fig. \ref{fig:q_dep}(c)). These behaviors are in contrast to the paramagnon at $q_\parallel$ = 0.33\,r.l.u., where the peak position remains unchanged and the spectral weight decreases. 

Apparently, the light-driven paramagnon variation exhibits a momentum dependence within the Brillouin zone. Figure~\ref{fig:q_dep}(d) shows the paramagnon dispersion along the $h$-direction before and after time zero, obtained by fitted paramagnon peak positions as a function of  $q_{\parallel}$ at $\Delta t = -2$\,ps and 0.25\,ps, respectively (see Supplementary Materials Fig. S5 for data and fitting). The paramagnon softening is most prominent near the zone center and diminishes when moving toward the zone boundary, resulting in a modulation of the paramagnon dispersion without changing the paramagnon bandwidth. Moreover, the light-induced change in paramagnon spectral weight is also momentum-dependent, as shown in Fig. \ref{fig:q_dep}(e). For small momenta, the pump increases the spectral weight due to a significant portion of light-driven paramagnons in the energy gain via anti-Stokes scattering (see Supplementary Material). Interestingly, the enhancement in spectral weight rapidly decreases with increasing momentum and eventually changes sign near $q_{\parallel}\sim 0.2$\,r.l.u., beyond which the overall spectral weight begins to decrease due to the pump. Altogether, our results, for the first time, demonstrate that the photoexcitation can modify the paramagnon dispersion and redistribute the spectral weight within the Brillouin zone.  

We remark that the dispersion modification does not necessarily imply that the underlying spin exchange interaction is modulated by the pump. Particularly, in the small $q_{\parallel}$ regime where the energy scale is also smaller, the paramagon peak position softening can be due to the substantial anti-Stokes fraction in the overall paramagnon spectral weight (see also Supplemental Material Fig. S2 and Fig. S4 \cite{Sup}). The fact that the paramagnon peak position remains unchanged near the zone boundary (\textit{i.e.} identical paramagnon dispersion bandwidth) thereby asserts no alteration in the spin-exchange coupling strength, at least along the $h$ direction in our pump-probe geometry.    

To contextualize our findings, we correlate the observed paramagnon evolution with quasiparticle dynamics in the photoexcited state \cite{Perfetti2007,Zonno2021}. Near time zero, photoexcitation generates hot electrons that rapidly form a pseudo-Fermi distribution with an effective electronic temperature. In this period, the paramagnon population increases, their dispersion is modulated, and spectral weight is redistributed in the Brillouin zone. The initial relaxation of hot electrons occurs by transferring energy to strongly coupled phonons on a timescale of a few hundred femtoseconds, matching the initial paramagnon recovery observed in our data. Thus, after photoexcitation, hot-electron relaxation and restoration of magnetic degrees of freedom proceed in tandem. Beyond a few picoseconds, the electronic, magnetic, and lattice subsystems reach a quasi-equilibrium and then recover more slowly.

In this context, a natural question is whether the light-induced paramagnon response can be described in terms of an effective “magnetic” temperature, similar to the typical description of quasiparticles. Although the broadening of the paramagnon peak and the increased paramagnon population resemble those of magnetic excitations at elevated temperatures, the altered dispersion at fixed bandwidth and the associated redistribution of spectral weight do not fully align with a simple heating picture. Our determinant quantum Monte Carlo (DQMC) simulations based on canonical ensemble\,\cite{Ding2024} show that, at high temperatures, the paramagnon energy at the zone boundary should also be reduced (see Supplemental Material Fig. S6), because thermally generated doublon–hole pairs, which move randomly in the copper-oxide plane, substantially lower the average energy required to flip a spin and decrease the spin correlation length. In contrast, in the photoexcited nonequilibrium state, the pump field can additionally drive mobile charge carriers through the antiferromagnetic background along the direction of the pump’s electric field, thereby scrambling spin orientations and perturbing the magnetic subsystem in a manner that cannot be captured solely by an effective temperature increase.

Since the spectral weight of a bosonic mode controls the strength of electron–boson coupling, any redistribution of paramagnon spectral weight in the Brillouin zone can transiently alter the coupling between AFM correlations and Fermi-surface quasiparticles. Specifically, our observation of a reduced spectral weight at large $q_{\parallel}$ implies a weaker coupling between electrons and magnetic fluctuation at higher momentum, likely including those near the ($\pi,\pi$) point. Consequently, this diminished electron–AFM coupling may be a key factor contributing to the filling-in of the pseudogap at the Fermi-surface “hot spots” in photo-excited NCCO \cite{Boschini2020}.

More broadly, the ultrafast modulation of dispersion and the generation of substantial populations of high-momentum magnetic modes offer a compelling pathway for magnonic devices \cite{Chumak2015}, which require efficient magnon generation and transport. Given that prototypical materials for magnonic applications—such as yttrium iron garnet (YIG)—are characterized by robust long-range magnetic order and intrinsically low magnetic damping, we expect that optically generated magnons in these systems will exhibit substantially longer lifetimes than the paramagnons observed in optimally doped NCCO. In summary, our results reveal that photoexcitation can reshape collective magnetic behavior. Looking ahead, especially when paired with tailored pump pulses, ultrafast excitation offers powerful new leverage for actively controlling the magnetic properties of materials. 
\\

\begin{acknowledgments}
The Experimental work was supported by the U.S. Department of Energy, Office of Basic Energy Sciences, Materials Sciences and Engineering Division under contract No.~DE-AC02-76SF00515. Theoretical simulations (J.H., S.D., and Y.W.) were supported by the U.S. Department of Energy, Office of Science, Basic Energy Sciences, under Early Career Award No.~DE-SC0024524. We acknowledge European XFEL in Schenefeld, Germany, for provision of X-ray free-electron laser beamtime at SCS under proposal number 3078 and would like to thank the staff for their assistance. D.J. gratefully acknowledges funding by the Alexander-von-Humboldt foundation. The simulation used resources of the National Energy Research Scientific Computing Center, a U.S. Department of Energy Office of Science User Facility located at Lawrence Berkeley National Laboratory, operated under Contract No.~DE-AC02-05CH11231. T.C.A. acknowledges funding from the Heisenberg Resonant Inelastic X-ray Scattering (hRIXS) Consortium.
\end{acknowledgments}

\bibliography{Reference}

\section*{Supplemental Material for Collective magnetic excitations in a Photo-excited Electron-doped Cuprate Superconductor}

\section*{Materials and methods}

\section*{Methods}
\label{sec:methods}

\subsection{Sample synthesis}
Single crystals of Nd$_{1-x}$Ce$_x$CuO$_4$ ($x = 0.15$, $T_c = 25$\,K) were grown using the floating zone method. The crystals were oriented with Laue. A fresh ab-plane surface was obtained by cleaving in the air right before the experiment.

\subsection{X-ray FEL configuration}
The X-ray FEL is operated at the standard pulse train mode with 10 trains per second, 1.1\,MHz repetition rate within each train of 360 pulses. The average X-ray pulse energy is 1000\,nJ with a pulse duration of 80\,fs. For the trRIXS measurement, 90\% of the X-ray pulse energy is attenuated to avoid sample heating, while still yielding a strong RIXS signal. The photon energy is tuned to the Cu $L$-edge for the RIXS measurements. The beam spot on the sample is set to 10 $\mu$m (V) x 150 $\mu$m (H). 

\subsection{Optical Penetration Depth of the 400\,nm Pump Pulses}
In order to estimate the penetration of the 400\,nm pump laser, we estimate the complex refractive index of NCCO at 15\% doping by measuring the angle-dependent reflectivity of 400\,nm light for $\sigma$- and $\pi$-polarizations.
We use the frequency-doubled output of a Ti:Sa oscillator at 80\,MHz with ~100\,fs pulse duration and $<$38\,mW incident power (0.5\,nJ pulse energy). NCCO crystals were cleaved and measured in air at room temperature. In this characterization, we estimate the spot size on the sample to be ~100\,$\mu$m. At normal incidence, this results in a fluence of 4.2\,nJ/cm$^2$ and a heating power of 335\,W/cm2, and we assume heating effects to be negligible.

The data and the fitting to determine the index of refraction at 400\,nm is shown in Supplemental Material Fig.~\ref{fig:SP_Experiment_Geometry} (b). The data is fit globally by applying the Fresnel equations to $\sigma$- and $\pi$-polarized data simultaneously.
\begin{align*}
R_{\pi} = |r_{\pi}^2|, r_{\pi} = \frac{n^2 \cos{\alpha_i}-\sqrt{n^2-\sin{\alpha_i}}}{n^2 \cos{\alpha_i}+\sqrt{n^2-\sin{\alpha_i}}}\\
R_{\sigma} = |r_{\sigma}^2|, r_{\sigma} = \frac{\cos{\alpha_i}-\sqrt{n^2-\sin{\alpha_i}}}{\cos{\alpha_i}+\sqrt{n^2-\sin{\alpha_i}}}
\end{align*}

Where $n$ is the complex index of fraction $n$=n + \textit{i}k. We obtain n=1.76(3) and k=0.59(2). With $\delta = \frac{\lambda}{4\pi k}$, we arrive at an optical penetration depth of the pump pulse of 400 nm to be 54 nm. 

\subsection{trRIXS configurations}
The trRIXS measurements were performed at the SCS instrument using the h-RIXS spectrometer \cite{Schlappa2025} with a scattering and pump-probe geometry shown in Supplemental Material Fig. \ref{fig:SP_Experiment_Geometry}(a). The scattering angle is set to be 125°. The samples were mounted on the 6-axis in-vacuum diffractometer and cooled to $\sim 20$ K. The RIXS data were obtained with incident $\pi$ polarization (in the scattering plane, high through-put configuration). The incident photon energy was tuned to the maximum of the Cu $L_3$-edge X-ray absorption spectrum at $\sim 931$ eV. The energy resolution was approximately $\Delta E \sim 120$ meV. The scattering plane coincided with the a-c plane of the crystal with a surface normal along the c-axis. The momentum transfer was calculated using the lattice constants of $a = b = 3.9$\,\AA~and $c = 12.1$\,\AA. The scattering geometry is optimized for the paramagnon measurement. Since the paramagnon is quasi-two-dimensional, the RIXS spectra presented here were plotted as a function of in-plane momentum only.  In our experiment, we rotate the sample under the collinear pump and probe beams such that theta ranges from 48 to 100 degrees, corresponding to a momentum transfer of q$_{parallel}$ ranging from -0.125 r.l.u to 0.33 r.l.u. Unless otherwise indicated, the RIXS spectra were normalized by the incident photon flux.

Pump pulses with a wavelength of 400 nm and a time duration of 50\,fs were used to drive the system. The data shown in the manuscript were taken with a pump fluence of 7 mJ/cm$^2$. The total time resolution, estimated to be ~150\,fs, is a convolution of X-ray FEL jitter, and the x-ray and pump pulse durations. The pump laser spot size was set to 120\,$\mu$m (V) x 300\,$\mu$m (H), which is substantially larger than the x-ray spot on the sample, ensuring homogeneous photon excitation within the probed region. The pump beam and x-ray beam were collinear before they reached the sample. The polarization of the pump laser was set to be perpendicular to the scattering plane, such that the relative orientation between the pump polarization and the crystalline axis of the sample remained unchanged when we rotated the sample angle for different momentum positions. The pump power was adjusted for different sample angles so that the pump fluence was kept fixed at 7\,mJ/cm$^2$.

We can estimate the 400\,nm pump excitation density within the XFEL probe beam footprint using our experimental parameters and the reflectivity data (Supplemental Material Fig.~\ref{fig:SP_Experiment_Geometry}(b)). Reflectivity measurements for the $\sigma$-polarized pump laser, averaged over the experimental sample angles, yielded a value of approximately 0.2. This corresponds to an absorption coefficient of 0.8 and a pump penetration depth of approximately 50\,nm (\textbf{Methods}, previous section). Based on these values, we estimate an absorption of approximately 0.6 photons per unit cell per pulse.

\subsection{Fitting of RIXS spectrum}
All the RIXS data were normalized to incident photo flux. For each time delay and momentum position, the zero-energy-loss position were determined by the fitted elastic peak position. The fitting model involve a Voigt function for the elastic peak with a full width half maximum of $\sim$ 0.12\,eV and a Gaussian-Lorentzian factor of $\sim$ 0.2, corresponding to the instrument resolution. The paramagnon and the acoustic plasmon were modeled using Gaussian functions. Since both paramagnon and plasmon are notably broader than the instrument resolution ($\sim0.12$\,eV), we did not deconvolve the instrument resolution from the fitted Gaussian peaks. In addition, a tail of Lorentzian is used to account for the for high energy background arisen from the $dd$ excitations.
We note that while it is generally assumed an odd function with respective to the zero energy, such as the anti-Lorentizan function or the damped harmonic oscillator function, to fit the magnetic excitation in RIXS, this assumption is only valid for static RIXS spectrum at low temperature. Because in the low temperature equilibrium state, the anti-Stoke scattering effect can be neglected, and thus the dynamical structure factor measured by RIXS can be approximated by the magnetic susceptibility, which possesses an odd parity. In the presence of a significant population of magnons, the dynamical structure factor can no longer be approximated as the susceptibility and a simple functional form is still not available. While in the thermal equilibrium condition, the dynamical structure can be related to susceptibility via dissipation-fluctuation theorem, such relation is not valid in the photo-excited nonequilibrium state. Therefore, we adopted the simplest model to fit the paramagnon and plasmon with a Gaussian by allowing the peak to cross the zero energy (see Supplemental Material Fig.~\ref{fig:SP_AntiStoke}). 

\subsection{Single-Band Hubbard Model}\label{Hubbard_model}
We use the single-band Hubbard model to simulate the NCCO system, whose Hamiltonian is
\begin{eqnarray}\label{eq:Hamiltonian}
{\mathcal{H}} &=& -t_\mathrm{h}\sum_{\langle \mathbf{i},\mathbf{j}\rangle\sigma} \left[c_{\mathbf{i}\sigma}^\dagger c_{\mathbf{j}, \sigma}  + \mathrm{H.c} \right]-t_\mathrm{h}^\prime \sum_{\langle\!\langle \mathbf{i},\mathbf{j}\rangle\!\rangle\sigma} \left[c_{\mathbf{i}\sigma}^\dagger c_{\mathbf{j}, \sigma} + \mathrm{H.c} \right]\nonumber\\
&&+ \, U \sum_{i} n_{i\uparrow}n_{i\downarrow} ,
\end{eqnarray}
where $c_{i\sigma}$ ($c_{i\sigma}^\dagger$) annihilates (creates) a valence electron and $n_{i\sigma} = c_{i\sigma}^\dagger c_{i\sigma}$ is the number operator. The valence electrons form a single band with nearest-neighbor hopping amplitude $t_\mathrm{h}$, next nearest-neighbor hopping amplitude $t_\mathrm{h}^\prime$, and on-site Coulomb repulsion $U$. The parameters are chosen as $U=8t_{\rm h}$, $t'_h=-0.3t_{\rm h}$, and $t_h = 400\,$meV, following Ref.~\onlinecite{jia2016using}. 

For pump-probe simulations, we employ a 12-site cluster to simulate pump-probe spectroscopes for a system with 16.7\% electron doping, the minimal doping level on this finite cluster with SU(2) symmetry. These simulations assume zero temperature and the ground-state wavefunctions $|\Psi_G\rangle$ are obtained using the parallel Arnoldi method\,\cite{lehoucq1998arpack,  jia2017paradeisos}. Equilibrium finite-temperature simulations are performed on a larger $6\times6$ cluster using determinant quantum Monte Carlo (DQMC) \cite{BSS1981,White1989}.

\subsection{Pump-Probe Simulations by ED}\label{pump_probe_method}
The light-matter interaction is described through the Peierls substitution $t_hc^\dagger_{\mathbf{i} \sigma} c_{\mathbf{j}\sigma}  \rightarrow t_h e^{i\int_{\mathbf{r}_\mathbf{i}}^{\mathbf{r}_\mathbf{j}} \mathbf{A}(t)\cdot d\mathbf{r}} c^\dagger_{\mathbf{i} \sigma} c_{\mathbf{j}\sigma}$. In the presence of the laser pulse, the vector potential $\mathbf{A}(t)$ gives time dependence to the many-body Hamiltonian in Eq.~\eqref{eq:Hamiltonian}, denoted as $\mathcal{H}(t)$. To mimic the pump-probe experiment, we simulate the pump pulse with an oscillatory Gaussian vector potential:
\begin{equation}
    \mathbf{A}(t)=A_0\hat{e}_{\rm pol} \exp\left[-\frac{(t-t_0)^2}{2\sigma^2}\right]\cos(\Omega t)\,,
\end{equation}
and fix the polarization as $\hat{e}_{\rm pol}=\hat x$ and the pump frequency as $\Omega = 3.1\,$eV.  
The time evolution of the wavefunction in the presence of a laser field $| \Psi_e(t+\delta t)\rangle  \approx e^{-i\mathcal{H}(t)\delta t}| \Psi_e(t)\rangle$ is determined through the Krylov subspace technique\,\cite{manmana2007strongly}.

To describe the evolution of paramagnon excitations, we simulate the nonequilibrium dynamical spin structure factor. It is obtained through the integral of two-time correlations
\begin{eqnarray}\label{eq:trSqw}
\mathcal{S}({q}, \omega, t) = \frac1{N}&\!\iiiint\!&  dt_1dt_2 e^{i\omega (t_2 - t_1) }  g(t_1; t)g(t_2; t) \nonumber\\
&&\times \langle\Psi(t_2)| \rho_{-\mathbf{q}}^{\rm(spin)}\rho_{\mathbf{q}}^{\rm(spin)} |\Psi(t_1) \rangle \,.
\end{eqnarray}
Here, the $g(\tau; t)$ is a probe shape function at time $t$, taken as Gaussian $\exp[-(\tau - t)^2 / 2\sigma_\text{pr}^2] / \sqrt{2\pi \sigma_\text{pr}}$ in this work. The momentum-representation spin excitation operator is defined as $\rho_{\mathbf{q}}^{\rm(spin)}=\sum_i (n_{i\uparrow} - n_{i\downarrow}) e^{ir_i \cdot \mathbf{q}}$.

We also present the trRIXS simulation in the SI, as a more faithful description of the experimental settings. By explicitly considering the X-ray scattering process, involving an intermediate core level (specifically the $2p$ orbitals of Cu in the $L$-edge RIXS), the full Hamiltonian reads as 
\begin{eqnarray}\label{eq:Hubbard}
	{\mathcal{H}}' &=&  {\mathcal{H}} + \sum_{\mathbf{i}\alpha\sigma} E_{\rm edge}(1-n^{(p)}_{\mathbf{i}\alpha\sigma}) - U_c \sum_{\mathbf{i}\alpha\sigma\sigma^\prime} c_{\mathbf{i}\sigma}^\dagger c_{\mathbf{i}\sigma}(1-n^{(p)}_{\mathbf{i}\alpha\sigma^\prime})\nonumber\\
        &&+\lambda\sum_{\substack{i\alpha\alpha^\prime\\ \sigma\sigma^\prime}}p_{\mathbf{i}\alpha\sigma}^\dagger \chi_{\alpha\alpha^\prime}^{\sigma\sigma^\prime}p_{\mathbf{i}\alpha^\prime\sigma^\prime}.
\end{eqnarray}
Here, the $p_{\mathbf{i} \alpha \sigma}$ ($p_{\mathbf{i} \alpha \sigma}^\dagger$) annihilates (creates) a core-level $2p_{\alpha}$ electron ($\alpha\! =\! x, y, z$) and $n^{(p)}_{\mathbf{i}\alpha\sigma}\!=\! p_{\mathbf{i}\alpha \sigma}^\dagger p_{\mathbf{i}\alpha \sigma}$ are the core-level electron density operators. The core-hole potential $U_c$ is fixed at 4$t_h$ and regarded identical for all 2$p$ orbitals\,\cite{jia2016using}. $E_{\rm edge}=938$\,eV represents the Cu $L$-edge absorption energy, i.e., the energy difference between the $3d$ and $2p$ orbitals without spin-orbit coupling. Finally, the spin-orbit coupling with the X-ray-induced core hole in the degenerate $2p$-orbitals is accounted by the last term in Eq.~\eqref{eq:Hubbard} with $\lambda=13$\,eV\,\cite{tsutsui2000resonant,kourtis2012exact}.

The trRIXS cross-section reads as\,\cite{wang2021x}
\begin{eqnarray}\label{eq:cross}
\mathcal{I}({q}, \omega_s, \omega_i, t) = &&\frac1{2\pi N}\!\iiiint\!  dt_1dt_2 dt_1'dt_2'e^{i\omega_i (t_2 - t_1) -i\omega_s(t_2' - t_1')}  \nonumber \\
&& \times g(t_1; t)g(t_2; t) l(t_1' - t_1) l(t_2' - t_2) \nonumber\\
&&\times \langle \hat{\mathcal{D}}^\dagger_{{q}_i{\varepsilon}_i}(t_2) 
\hat{\mathcal{D}}_{{q}_s{\varepsilon}_s}(t_2') 
\hat{\mathcal{D}}^\dagger_{{q}_s{\varepsilon}_s}(t_1') 
\hat{\mathcal{D}}_{{q}_i{\varepsilon}_i}(t_1) 
 \rangle 
\end{eqnarray}
where $q=\mathbf{q}_i-\mathbf{q}_s$ ($\omega=\omega_i-\omega_s$) is the momentum (energy) transfer between incident and scattered photons, and $l(t)\! =\!  e^{- t/\tau_{\rm core}}\theta(\tau)$ the core-hole decay lifetime. For a direct transition, the dipole operator reads as $\mathcal{D}_{\mathbf{q}\varepsilon}=\sum_{\mathbf{i}\alpha\sigma}e^{-i\mathbf{q}\cdot \mathbf{r}_\mathbf{i}}(M_{\alpha\varepsilon} c_{\mathbf{i}\sigma}^\dagger p_{\mathbf{i}\alpha\sigma}  + h.c.)$. We further select $\pi-\sigma$ polarizations for the incident and scattering photons to maximize contributions from spin-flip processes\,\cite{ament2009theoretical,Braicovich2010magnetic,Haverkort2010theory,jia2016using,Robarts2021dynamical}.



\clearpage
\newpage 
\onecolumngrid

\setcounter{page}{1}
\setcounter{figure}{0}


\begin{figure*}[t]
    \centering
    \includegraphics[]{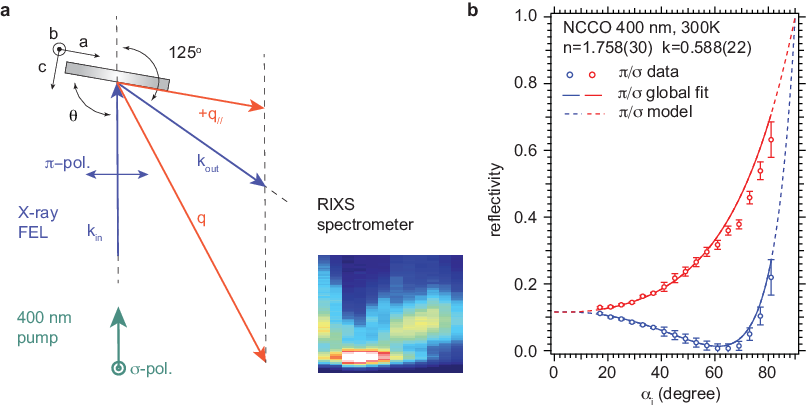}
    \caption{\textbf{Scattering and Pump-probe Geometry and NCCO reflectivity to 400 nm.} (a) $k_{in}$ and $k_{out}$ represent the momenta of the incident and outgoing x-ray, respectively. The scattering angle is 125$^{\circ}$. The ac plane of the crystal is in the scattering plane, as sketched.  $q$ and $q_{\parallel}$ represent the magnitude of the momentum transfer and the projected in-plane momentum along the a-axis (i.e. $h$-direction). The $q_{\parallel}$ is defined to be positive when $\theta > 62.5^{\circ}$ (\textit{i.e.} on the branch of grazing x-ray emission). The polarization of the incident x-ray lies in the scattering plane (\textit{i.e.} $\pi$ polarization). The 400 nm pump is collinear with the incident x-ray with a $\sigma$ polarization (b) Angle-dependent reflectivity for $\sigma$- and $\pi$-polarized 400 nm light. $\alpha_i$ denotes the angle-of-incidence as measured from the surface normal. Global fits of the Fresnel equations are shown as solid lines. Dashed lines are extrapolation of the fits.}
    \label{fig:SP_Experiment_Geometry}
\end{figure*}

\begin{figure*}[t]
    \centering
    \includegraphics[]{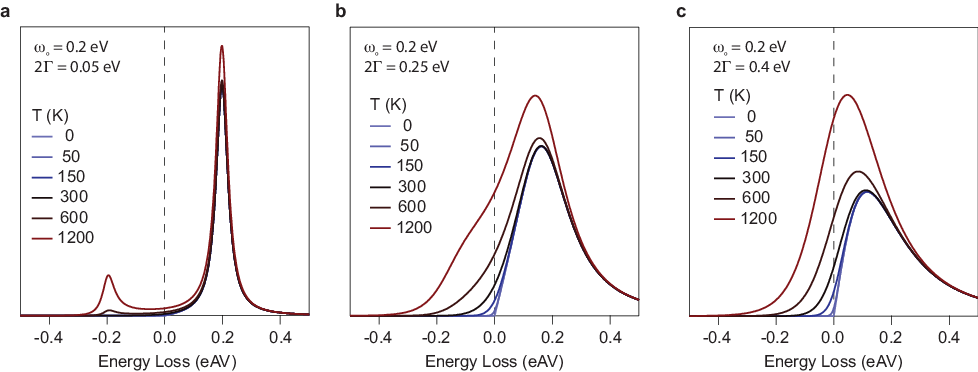}
    \caption{\textbf{Illustration of energy gain spectrum in thermal equilibrium condition.} When a finite population of bosonic modes exists, anti-Stokes scattering becomes detectable, resulting in an energy gain spectral weight in the dynamical structure factor ($S(q,\omega)$), which can be measured using inelastic neutron and X-ray scattering experiments. In the thermal equilibrium condition, modes are excited by thermal energy with a population following the Bose-Einstein statistics. Using the dissipation-fluctuation theorem, $S(q,\omega) = (1 + n(\omega, T)) \cdot \chi(q,\omega)''$, where $n(\omega, T)$ and $\chi(q,\omega)''$ are the Bose-Einstein distribution function and dynamical susceptibility $\chi(q,\omega)''$, respectively. 
    In this simulation, we use the damped harmonic oscillator function to generate $\chi(q,\omega)''$ for three different scenarios: a sharp mode ((a), $\omega > 2\Gamma$)), a damped mode ($\omega \sim 2\Gamma$), and an overdamped mode ($\omega < 2\Gamma$). Only in the situation of a sharp mode, a distinct peak can be resolved in the energy gain spectrum, which is the commonly known anti-Stoke peak. In the situations of a broad mode ((b), (c)), the anti-Stoke peak is not resolvable, instead the $S(q,\omega)$ manifests as a broadened peak with significant spectral weight in the negative energy loss (energy gain). 
    Regarding the case of photoexcitation, the dissipation-fluctuation theorem cannot be applied due to the non-equilibrium nature near time zero. Nevertheless, we expect a qualitative similarity for the spectral lineshape in the presence of photo-excited mode population. Since the paramagnon and plasmon are heavily damped in the optimally-doped electron-doped cuprates, we believe that a Gaussian function is a realistic and simplest model to fit the time dependence of the paramagnon and plasmon.} 
    \label{fig:SP_AntiStoke}
\end{figure*}

\begin{figure*}[t]
    \centering
    \includegraphics[width = 0.9\columnwidth]{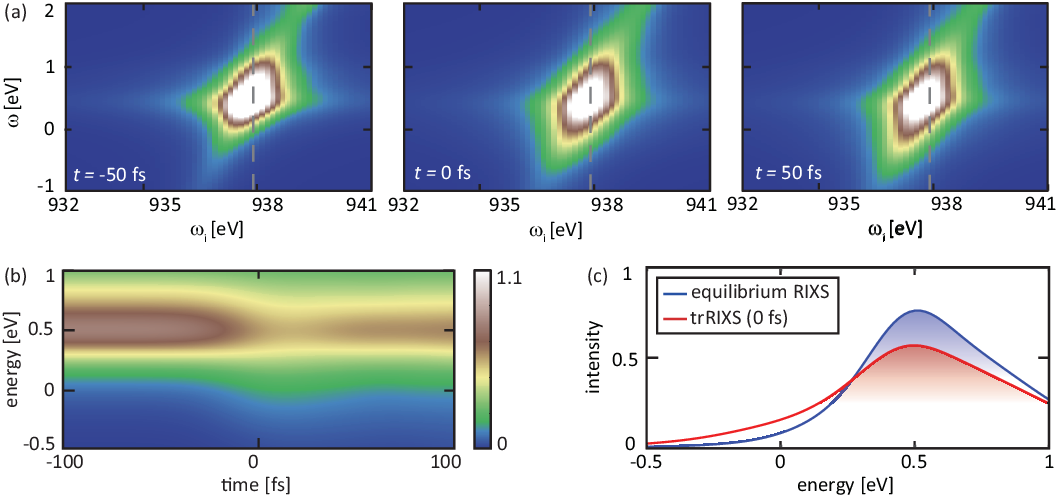}
    \caption{\textbf{Exact Diagonalization time-resolved RIXS calculations} a. Snapshots of the trRIXS spectra at $q = (0.33, 0)$ (2$\pi$/a) for the system before the pump, at the pump center, and after its departure. The dashed line indicates the resonance incident energy $\omega_i$.  b. Time evolution of the RIXS scattering cross section by fixing the incident energy at $\omega_i = 5.12$ eV.  c. Spectral cuts of trRIXS before and at the center of the pump.}
    \label{fig:SP_theory_trRIXS}
\end{figure*}

\begin{figure*}[t]
    \centering
    \includegraphics[]{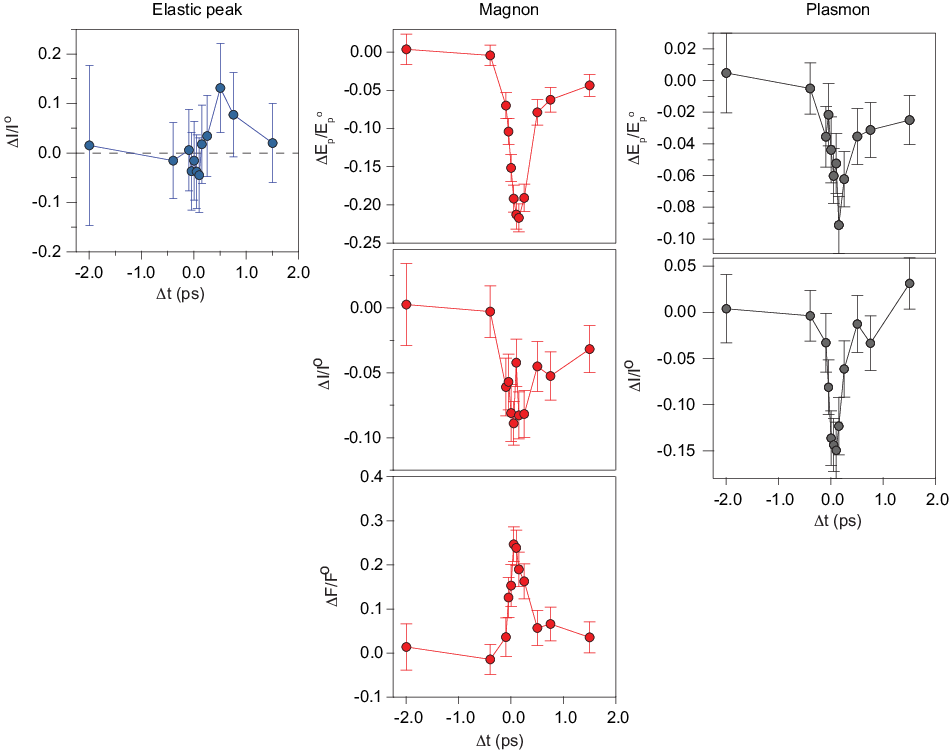}
    \caption{\textbf{Fitted parameters of $q_{\parallel}=-0.125$\,r.l.u. data.} All fitted parameters for the $q_{\parallel}=-0.125$\,r.l.u. data shown in Fig. 4(a,b). We found that if the plasmon width was used as a fitting parameter for the spectra after time zero, it decreases, which we believe is not physical. Thus, the width of the plasmon was first determined in the spectrum before time zero ($\Delta t = -2.0$ ps); then it was kept fixed for spectra at later time delays.}
    \label{fig:SP_q0p125}
\end{figure*}

\begin{figure*}[t]
    \centering
    \includegraphics[]{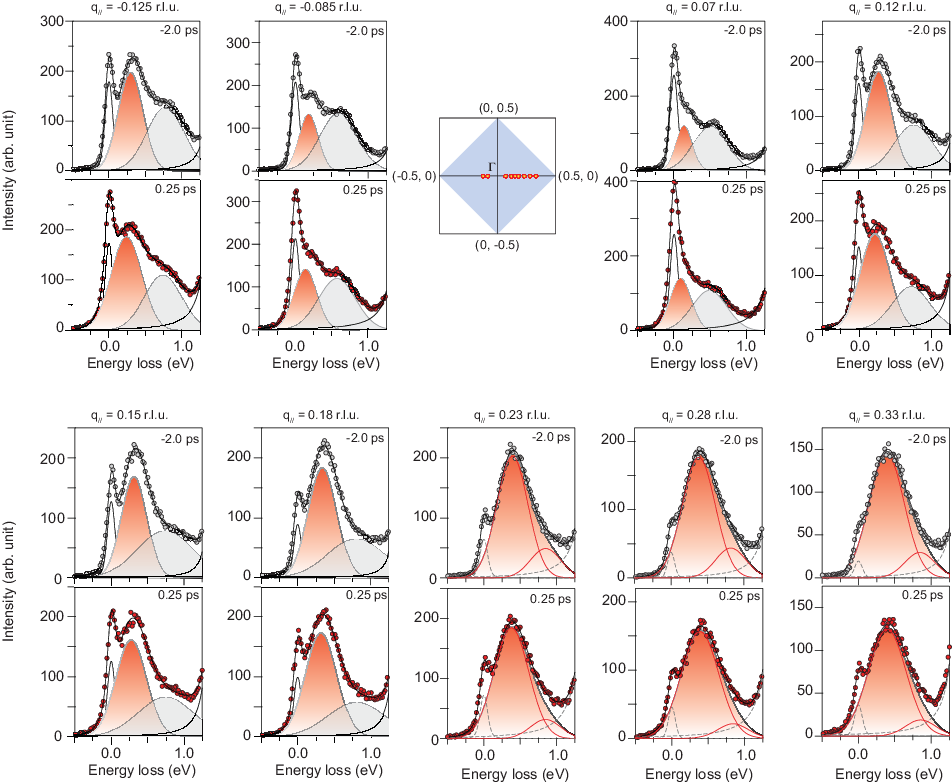}
    \caption{\textbf{Raw RIXS data and fitting.} Raw data (markers) used for momentum dependence analysis in Figs. 2 and 4(d,e) are shown here. The fitted curves generated from the minimal model described in \textbf{Methods} are superimposed as black curves. The red and black shaded peaks represent the fitted paramagnon and plasmon excitations, respectively. We note that for $q_{\parallel} \ge 0.23$\,r.l.u., the paramagnon cannot be described by a single Gaussain peak to the slight asymmetry in the RIXS lineshape. To account for spectral weight of the entire paramagnon peak, an additional Gaussain peak is included at high energy.}
    \label{fig:SP_All_q_dep_data}
\end{figure*}


\begin{figure*}[t]
    \centering
    \includegraphics[]{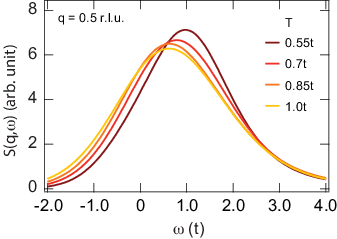}
    \caption{\textbf{Temperature evolution of $S(q,\omega)$ via DQMC.} The calculated $S(q,\omega)$ at the zone boundary (q = 0.5\,r.l.u.). At elevated temperatures, paramagnon peak energy exhibit apparent softening due to the generation of ``doublon-hole" pairs, which substantially lower the average energy required to flip a spin and decrease the spin correlation length. The unit of the energy is the hopping integral $t$.}
    \label{fig:SP_DQMC}
\end{figure*}

\end{document}